\documentclass[showpacs,graphicx,floatfix,twocolumn,prl]{revtex4}
\usepackage{amssymb}
\usepackage{epsf,bm}
\usepackage{amsmath}
\usepackage{graphicx}
\newcommand{\FigWidth}{\columnwidth}
\begin{document}

\tighten

\title{Raman resonance in spin $S$ two-leg ladder systems}

\author{A. Donkov$^1$ and A. V. Chubukov$^{1,2}$}
\affiliation{$^1$ 
Department of Physics, University of Wisconsin, Madison, WI 53706\\
$^2$Department of Physics and Condensed Matter Theory Center,
University of Maryland, College Park, MD 20742-4111.}
\date{\today}

\begin{abstract}
We argue that the Raman intensity in a spin $S$ two-leg spin-ladder
 has a pseudo-resonance peak, whose width is very small at large $S$.
 The pseudo-resonance
 originates from the existence of a local minimum in the magnon 
excitation spectrum, and  is located slightly  below twice the 
 magnon energy at the minimum. The physics behind the peak is similar to 
the excitonic scenario for the neutron and Raman resonances in 
a $d-$wave superconductor.
 \end{abstract}


\maketitle

Recently, there has been a considerable 
 experimental progress in Raman studies of two-leg spin-ladder materials 
$(Sr,La)_{14} Ca_x Cu_{24} O_{41}$ (Ref~\cite{girsh,sugai}) and 
$SrCu_2O_3$~\cite{Thomsen}.
The most intriguing experimental result is the discovery of a sharp
 peak in the Raman intensity of $Sr_{14}Cu_{24}O_{41}$ 
 at a frequency near $3000 cm^{-1}$ (about $400 meV$). The peak exists for polarizations of incoming and outgoing light both along  and across the ladders
 ($xx$ and $yy$, respectively), and its width for xx
 polarization is 
 around $130 cm^{-1}$ which is nearly 10 times smaller 
than the width of the two-magnon peak in 2D antiferromagnets~\cite{girsh}. 

The discovery of the peak stimulated the search for  possible resonance-like 
 features in the two-magnon Raman profile $R(\omega)$ of spin-ladder systems
~\cite{uhrig,Trebst,uhrig_2,FreitasSingh}. 
Previous studies of $S=1/2$ ladders 
 didn't find the resonance and suggested more complex explanation of the 
 sharp peak in $R(\omega)$~\cite{uhrig,uhrig_2}.
 We argue in this paper that  the Raman intensity in a spin $S$ 
two-leg ladder  possesses a pseudo-resonance, whose origin is similar to
 the origin of the 
neutron resonance and $B_{1g}$ Raman pseudo-resonance in high $T_c$ 
superconductors~\cite{neutrons,cmb}.
 The intrinsic width of the Raman resonance is small in $1/S$  
 for $S \gg 1$, but increases as $S$ decreases. 
We discuss whether the sharp Raman 
peak observed in $S=1/2$ material 
$Sr_{14}Cu_{24}O_{41}$  may be this resonance.

\begin{figure}
\includegraphics[clip=true,width=\FigWidth,height=1.3in]{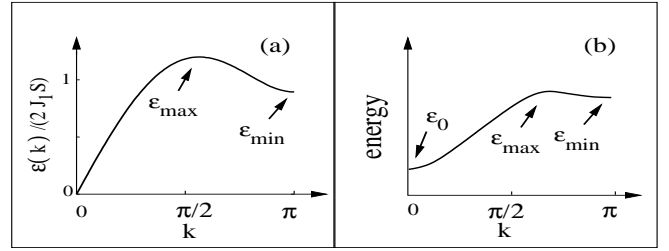}
\caption{The magnon excitation spectrum in a two-leg ladder.
 Left panel, the quasiclassical case $S \gg 1$, $J_2/J_1 = 0.4$. Right
 panel - schematic $\epsilon_k$ for $S=1/2$~\protect\cite{uhrig}. 
In both cases, the spectrum has a local minimum at $k=\pi$ and a local maximum 
 at some smaller $k$. In the quasiclassical case, the excitation
 spectrum is gapless at $k=0$. For $S=1/2$ ladder, magnon states near
 $k=0$ are gapped due to the Haldane effect. Note that 
our momenta are shifted by $\pi$ compared to \protect\cite{uhrig}.
The local minimum disappears at $J_2/J_1 = 2$ both in the quasiclassical case and for $S=1/2$~\protect\cite{Trebst}
} 
\label{fig_0}
\end{figure}

\begin{figure}
\includegraphics[clip=true,width=\FigWidth,height=1.4in]{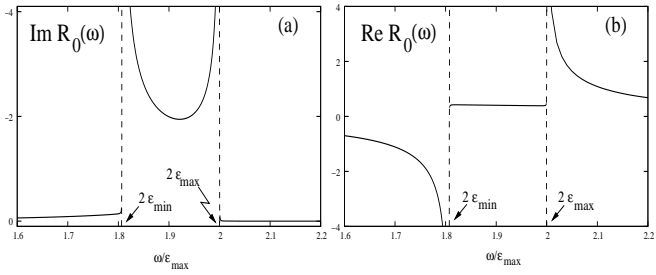}
\caption{The behavior of the Raman bubble $R_0 (\omega)$ (in units of $J_1$
 for non-interacting magnons. We used $J_2/J_1 =0.8$ for illustrative purposes.
 Near $2\epsilon_{max}$ and 
 $2 \epsilon_{min} \approx 1.8 \epsilon_{max}$,
 the polarization bubble diverges as a square-root
due to  singularities in the magnon density of states.
Observe that while $Im R_0 (\omega)$ is negative for all frequencies,
$Re R_0 (\omega)$ is positive above $2\epsilon_{max}$,
 but negative below $2\epsilon_{min}$. This last behavior leads to a resonance in the Raman intensity below $2\epsilon_{min}$. In between $1.8 \epsilon_{max}$ and $2\epsilon_{max}$, $Re R_0 (\omega)$ is nonzero, but very small.} 
\label{fig_1} 
\end{figure}

\begin{figure}
\includegraphics[clip=true,width=\FigWidth,height=1.4in]{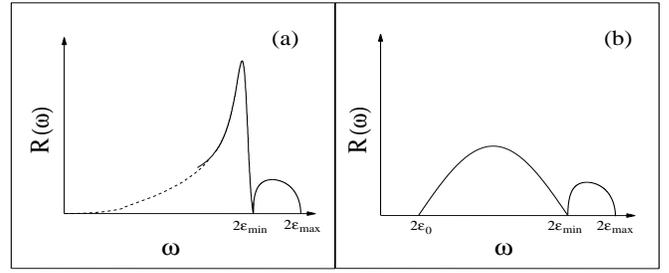}
\caption{The theoretical behavior of the full Raman intensity 
 $R(\omega)$ given by Eq. (\protect\ref{1}). Left panel, the
 quasiclassical case formally extended to $S = 1/2$. Solid line is the
 result of numerical calculations valid at large $\omega$, near 
 twice minimum and maximum of the magnon spectrum. 
 Dashed line is the expected behaviour for small $\omega.$ 
We used $J_2/J_1 =0.8$, as in Fig.\protect\ref{fig_1}.
Right panel, the expected behavior of $R(\omega)$ for $S=1/2$.  
The width of the pseudo-resonance below $2\epsilon_{min}$, obtained in numerical simulations for $S=1/2$ ladder~\protect\cite{uhrig,uhrig_2},
 is larger than in the quasiclassical analysis extended to $S=1/2$.}
\label{fig_2}
\end{figure}

The  pseudo-resonance in a two-leg ladder 
 emerges because of the existence of 
 a local minimum in the  magnon spectrum. 
 The magnetic excitations in a ladder are
 well described by  Heisenberg interactions between spins along the
 legs of the ladder ($J_1$) and along the rungs of the ladder ($J_2$). In the quasiclassical (large $S$) approximation, the excitation spectrum consists of two branches $\epsilon_k$ and $\epsilon_{k+\pi}$ where 
\begin{equation}
\epsilon_k = 4 J_1 S |\sin (k/2)| \sqrt{\cos^2 (k/2) + J_2/2J_1}.
\label{new_3}
\end{equation}
 The spectrum $\epsilon_k$ is gapless at $k=0$, and  for $J_2 < 2J_1$,
 which we only consider, it reaches
 a maximum $\epsilon_{max} = S (2J_1 + J_2)$ at some finite $k$, then falls 
down and reaches a minimum $\epsilon_{min} = S \sqrt{8 J_1 J_2}$ 
 at $k=\pi$ (see  Fig.\ref{fig_0}a). 
At a finite $S$, Haldane effect~\cite{haldane} 
 produces a gap in $\epsilon_k$  at $k=0$~\cite{uhrig}, however, 
 the minimum at $k=\pi$ survives (see  Fig.\ref{fig_0}b).
We verified that the effects on the Raman profile from $\epsilon_k$ and $\epsilon_{k+\pi}$ are additive, so below we only discuss the dispersion branch
 $\epsilon_k$
We first present the summary of the results and then discuss the computations.
Following previous studies of $2D$ systems~\cite{singh,cf}, we assume that 
 the Raman intensity $R(\omega)$ in two-leg ladders is reasonably well 
approximated by the RPA expression~\cite{rpa} 
\begin{equation}
 R (\omega) = 
 -\frac{Im
     R_0}{\left( 1+ a Re R_0 \right)^2+ \left( a Im R_0 \right)^2},
\label{1}
\end{equation}
where $R_0 (\omega)$ is the polarization bubble of free fermions
 with Raman vertices, and positive $a$ accounts for
  magnon-magnon interaction. 
Near $\epsilon_{min}$ and $\epsilon_{max}$, excitation spectrum is flat, 
  and the magnon density of states diverges. This
 leads to square-root singularities in $R_0 (\omega)$ near 
  $2\epsilon_{min}$ and $2\epsilon_{max}$. Near these two singularities,
\begin{equation}
R_0 (\omega) =- \frac{A}{\sqrt{2 \epsilon_{min} - \omega -i\delta}}~~
R_0 (\omega) =\frac{A^*}{\sqrt{\omega - 2 \epsilon_{max} + i\delta}},
\label{new_1}
\end{equation}
where $A$ and 
$A^*$ are positive. 
The imaginary part of $R_0 (\omega)$ diverges  
 at  approaching $2\epsilon_{min}$ from above and at approaching 
 $2\epsilon_{max}$ from below  (see Fig. \ref{fig_1}).
One can easily make sure that in both cases
 $Im R_0$ is negative (and Raman intensity is positive). 
Meanwhile, $Re R_0 (\omega)$ is positive above $2\epsilon_{max}$,
 and negative below $2\epsilon_{min}$  (see Fig. \ref{fig_1}).
 A positive $Re R_0 (\omega)$ at $\omega > 2\epsilon_{max}$ implies that
 above $2\epsilon_{max}$,  $1 + a Re R_0 (\omega)$ in Eq. (\ref{1}) is
 far from zero, 
 i.e., there is no antibound state in the Raman profile. This is 
 consistent with previous studies of the Raman profile in 2D 
 Heisenberg systems~\cite{cf}.  At the same time, below 
 $2\epsilon_{max}$, $ Re R_0 (\omega) <0$, and therefore 
there exists a frequency $\omega_{res}$ 
 at which  $1 + a Re R_0 (\omega_{res}) =0$, and the Raman intensity 
 $R(\omega)$ is strongly enhanced. If
$\epsilon_{min}$ was a true minimum of the excitation spectrum, 
the full $R(\omega)$ given by Eq. (\ref{1}) would develop a truly 
$\delta-$functional resonance peak at $\omega = \omega_{res}$. 
In reality,  there are magnon states 
 below $\epsilon_{min}$ (see Fig.\ref{fig_1}), and 
$Im R_0$ remains finite, albeit small below $2\epsilon_{min}$.
 In this situation,  the full 
Raman intensity acquires only a peak at $\omega = \omega_0$.
  The width of the peak scales as $1/S$ at large $S$,
 but is $O(1)$ for $S=1/2$ (see Fig.\ref{fig_2}).

The physics that we just described is very
 similar to the excitonic scenario
 for the resonance in the neutron scattering ~\cite{neutrons} and 
 the pseudo-resonance in $B_{1g}$ Raman intensity~\cite{cmb} in the cuprates.
 Like in cuprates, the resonance in the two-leg ladders
 is the combination of the two effects: the presence of the  gap
 in a single particle excitation spectrum, and the attractive residual interaction between quasiparticles. The attraction leads to a formation of a two-particle bound state below twice the gap, which shows up as a
  resonance in $R(\omega)$. The finite intrinsic width of the Raman 
resonance in a ladder is due to the fact that Raman intensity is a $q=0$ probe, and it includes a contribution from the states for which $Im R_0 (\omega)$
 is finite  below $2\epsilon_{min}$.

In $Sr_{14}Cu_{24}O_{41}$, 
the sharp  peak in the Raman profile has been detected around  $3000 cm^{-1}$
~\cite{girsh,sugai}. The measured maximum frequency $2 \epsilon_{max}$ is near $4000 cm^{-1}$.
 Whether or not this peak is our pseudo-resonance 
 depends on whether this material is truly described by a $S=1/2$ ladder,
 or Haldane effect is suppressed by $3D$ couplings, and the 
 form of the excitation spectrum is similar to the quasiclassical expression.
In the first case, numerical studies indicate~\cite{uhrig,uhrig_2} that 
the pseudo-resonance is too broad to account for the data.
 However, if the quasiclassical description is valid
 down to  $S=1/2$, the pseudo-resonance 
 below $2 \epsilon_{min} = 2 \sqrt{2 J_1 J_2}$ is quite sharp (Fig.\ref{fig_2}a) and the profile of $R(\omega)$ is consistent with the data.
Matching the peak position and the location of the upper edge for 
 $R(\omega)$ by quasiclassical formulas yields 
$J_1 \sim 1600 cm^{-1}$ ($200 meV$) and 
 $J_2/J_1 \sim 0.4-0.5$. 
This value of $J_1$ is comparable to  $J \sim 100 meV$
 in the 2D cuprates ~\cite{2d_exchange}, the value of $J_2/J_1$ roughly agrees with other estimates~\cite{girsh,FreitasSingh}. 

In the rest of the paper we present the details of 
our derivation of the Raman intensity.
Our point of departure is the Hubbard Hamiltonian for two chains
\begin{eqnarray}
H &=& -t \sum_{<i,j>} (c^\dagger_{i,\sigma} c_{j, \sigma} +d^\dagger_{i,\sigma} d_{j, \sigma}) -t' \sum_i
c^\dagger_{i,\sigma} d_{i, \sigma}+h.c. \nonumber \\
&& + U \sum_i \left( n_{i,\uparrow}^c n_{i, \downarrow}^c + n_{i,\uparrow}^d n_{i, \downarrow}^d \right).
\label{a1}
\end{eqnarray}
In the quasiclassical case, we introduce antiferromagnetic long-range order
 at $Q= (\pi,\pi)$  via
 $<\sum_k c^\dagger_{k+Q, \uparrow} c_{k, \uparrow}> = \alpha$, and $
< \sum_k d^\dagger_{k+Q, \uparrow} d_{k, \uparrow}> = -\alpha$.
Decoupling the Hubbard $U$ term using these relations, diagonalizing the
 resulting quadratic Hamiltonian, and introducing
 new valence ($b,f$) and conduction ($a,e$) electrons instead of
$c_k, c_{k+Q}, d_k, d_{k+Q}$,  we obtain
$$
H = {\textstyle{\sum^{ '}}}{\rm E}_k ( a^\dagger_{k,\sigma} a_{k, \sigma}-b^\dagger_{k,\sigma} b_{k, \sigma})
+\tilde{{ \rm E}}_k ( e^\dagger_{k,\sigma} e_{k, \sigma}-f^\dagger_{k,\sigma} f_{k, \sigma}),
$$
where ${\rm E}_k = \sqrt{ (2t \cos k + t')^2 + \Delta^2} \;\;
\tilde{{\rm E}}_k = \sqrt{ (2t \cos k - t')^2 + \Delta^2}$, $\Delta = U \alpha$,  and prime indicates that the summation goes over magnetic Brillouin zone. The
 self-consistency equation on $\alpha$ yields $\alpha = f(t/U, t^\prime/U)$,
 where $f(0,0) = 1/2$~\cite{swz}. 
 
The two-magnon Raman profile is observed in 
near-resonant Raman scattering regime, where
 the light couples to electrons predominantly
 via ${\bf j} {\bf A}$ term, where ${\bf A}$ is the vector potential of light, and ${\bf j}$ is electron current with the components 
$ j_x = {\textstyle{\displaystyle{\sum_{
        k,\sigma}}^{'}}}
2 t \sin(k) \left[ a^\dagger_{k, \sigma}  b_{k, \sigma} +
  e^\dagger_{k, \sigma}  f_{k, \sigma} +  h.c. \right], ~~
 j_y = {\textstyle{\displaystyle{\sum_{
        k,\sigma}}^{'}}}
 i t' \left[ a^\dagger_{k, \sigma}  f_{k, \sigma} -
  f^\dagger_{k, \sigma}  a_{k, \sigma} +  b^\dagger_{k, \sigma}  e_{k, \sigma} -
  e^\dagger_{k, \sigma}  b_{k, \sigma} \right]$,
where $x$ and $y$ are the directions along the chains and transverse to the chains, respectively.

\begin{figure}
\includegraphics[clip=true,width=\FigWidth,height=1.3in]{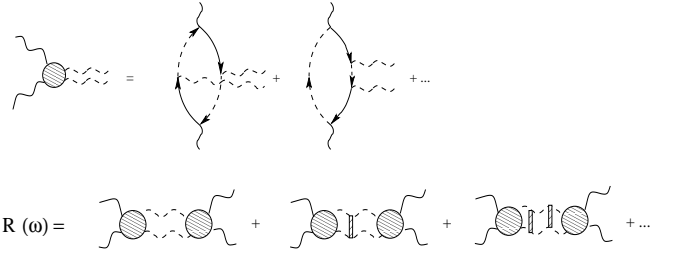}
\caption{a) Diagrammatic representation of the Raman vertex for the interaction between light and magnons. Solid wavy lines represent light, solid and dashed straight lines represent conduction and valence electrons, and dashed wavy lines represent magnons. There are other diagrams (not shown) whose role is to cancel 
 parasitic contributions from these two diagrams~\protect\cite{cf}. 
 b) Diagrammatic derivation of Eq. (\protect \ref{1}). Shaded rectangles represent $V(k,l)$.}
\label{fig_3}
\end{figure}

Other elements of two-magnon Raman scattering are the
 electron-magnon coupling, the magnon propagator, 
and  the magnon-magnon interaction. They all are obtained in a straightforward 
 manner from Eq. (\ref{a1}) by taking the large $U$ limit, extending the Hubbard model to large $S$, and computing
 the spin susceptibilities in the RPA approximation,
 which becomes exact in the quasiclassical case~\cite{swz,ss}. 
The poles of the spin susceptibilities determine
 magnon dispersion, and the effective electron-magnon Hamiltonian is  obtained
  by summing up RPA series of particle-hole renormalizations of the 
Hubbard $U$. Once this interaction is known, one can straightforwardly 
obtain the effective vertex $M_R (k, \omega)$ for the 
interaction between light 
and two magnons with momenta $k$ and $-k$ and total frequency $\omega$
 (see Ref. ~\cite{cf} and Fig.\ref{fig_3}).
Carrying out rather cumbersome calculations of the
 susceptibilities and vertices, we obtain at large $U$~\cite{comm}  
\begin{equation}
M_R (k) =  \frac{ 4 J_1 J_2 S (\cos(k)-1) }{2 \epsilon (k)}~ \left({\bf e}_{ix}  {\bf e}^\ast_{fx} - {\bf e}_{iy}{\bf e}^\ast_{fy}\right),
\label{a3}
\end{equation}
where ${\bf e}$ are unit vectors, 
 we introduced $J_1 = \frac{4 t^2}{4 S^2 U}, J_2 = \frac{4 t'^2}{4 S^2 U}$, and $\epsilon (k)$ is given by Eq. (\ref{new_3}).

The last element required for the computation of two-magnon Raman scattering 
 is the magnon-magnon interaction. To derive it, we note that at large $U$,
 the magnetic properties of the Hubbard model are adequately described
 by the effective Heisenberg Hamiltonian $H = \sum_{r,\mu} \left( 
J_1  {\bf S}_r {\bf S}_{r+\mu_x} + J_2  {\bf S}_r {\bf S}_{r+\mu_y}\right)$. 
 Applying the Holstein-Primakoff 
 transformation to this Hamiltonian, and diagonalizing the quadratic form, 
 we reproduce the magnon dispersion and also obtain the four-magnon interaction vertex. In the quasiclassical approximation, only the 
term  $H_{int} = - V(k,l) e^\dagger_k e^\dagger_{-k} e_l
e_{-l}$ with two creation and two annihilation boson operators $e$ is relevant~\cite{cf}.
For $V(k,l)$, we obtained
\begin{eqnarray}
&&V(k,l) = \frac{1}{4} \left[ J_2 + J_1 (\cos(k-l) + \cos(k+l)) \right]
( \mu_k^2 \mu_l^2 \nonumber \\
&& + \lambda_k^2 \lambda_l^2) +\left[ J_2 + J_1 (1 + \cos(k-l)) \right]  \mu_k \mu_l \lambda_k \lambda_l \nonumber  \\
&&-\frac{1}{4} \left[3 J_2 / 2 + J_1 (\cos(k) + 2 \cos(l)) \right]
 \mu_k \lambda_k \left( \mu_l^2 + \lambda_l^2 \right) \nonumber \\
&& -\frac{1}{4} \left[3 J_2 / 2 + J_1 (\cos(l) + 2 \cos(k)) \right]
\mu_l \lambda_l \left( \mu_k^2 + \lambda_k^2 \right).
\label{a5}
\end{eqnarray}
The coherence factors $\mu_k$ and $\lambda_k$ are given by
\begin{eqnarray}
\mu_k &=& \frac{1}{\sqrt{2}} \sqrt{\frac{2 J_1 S + J_2
    S}{\epsilon (k)}+1}, \nonumber \\
\lambda_k &=&\frac{1}{\sqrt{2}} \frac{2 J_1 S \cos(k)+J_2 S }{\left| 2 J_1 S \cos(k)+J_2 S
    \right|}\sqrt{\frac{2 J_1 S + J_2 S}{\epsilon (k)}-1}. 
\label{a6}
\end{eqnarray}

With these results at hand, we 
 can now compute the Raman intensity $R(\omega)$.
Without magnon-magnon interaction, the Raman intensity 
$R (\omega) = - Im R_0 (\omega)$, where
\begin{equation}
 R_0 (\omega) = \frac{i}{2 \pi} \sum_k \int
 d\omega^\prime
 M^2_R (k) G_{k,\omega^\prime} G_{-k, \omega - \omega^\prime},
\label{new_2}
\end{equation}
 and
$G_{k, \omega} = 1/(\omega -\epsilon (k) + i \delta {\text sgn} \omega)$ is a 
 magnon propagator. Substituting the result for $M_R (k)$ from (\ref{a3}) and evaluating the integral over $\omega^\prime$, we obtain 
\begin{equation}
R_0(\omega) = (2 J_1 J_2 S)^2 \sum_k \left[\frac{1 - \cos(k)}{\epsilon (k)}\right]^2 ~\frac{1}{\omega  - 2 \epsilon (k) + i \delta}.
\label{a7}
\end{equation}
Like we said, the magnon dispersion $\epsilon (k)$, Eq. (\ref{new_3}), 
 has a maximum $\epsilon_{max} =  2 J_1 S (1 + J_2/2 J_1)$ at $k = k_0 = 
\arccos (-J_2/(2J_1))$,
 and a  minimum $\epsilon_{min} = S \sqrt{8 J_1 J_2}$ at $k=\pi$.
Near the maximum and the minimum of $\epsilon (k)$, the magnon density of states $N(\epsilon) = d k/d\epsilon (k)$ diverges as a square-root of a distance to either $\epsilon_{min}$ or $\epsilon_{max}$. 
Replacing $\int dk $ by $\int N(\epsilon) d \epsilon$, we 
reproduce Eq. (\ref{new_1}).  Below $2 \epsilon_{min}$, $R_0 (\omega)$ is finite because of low-energy magnon states at small $k$, but is reduced 
 due to the factor $[(1- \cos k)/\epsilon (k)]^2 \propto k^2$.
 We found from (\ref{a7})
 that as $\omega$ approaches $2\epsilon_{min}$ from below, $Im R_0(\omega)$ tends to constant value. It then jumps to infinity at $\omega = 2\epsilon_{min} +0$, and  decays as $1/\sqrt{\omega -2 \epsilon_{min}}$ at larger frequencies(see Fig. \ref{fig_1}).

When magnon-magnon interaction is included, the full $R(\omega)$ is given
 by Eq. (\ref{1}), where $a = V(k,l)/(M_R (k) M_R (l))^{1/2}$ (see Fig. \ref{fig_3}b). Near  the two points where
 the density of states diverges, the interaction $V(k,l)$ from (\ref{a5}), (\ref{a6}) can be approximated by $V(\pi,\pi) =  (J_1 + J_2/2)/2$, and $V (k_0,k_0) =  J_2 (1 + J_2/2J_1)/4$. Using these forms, we obtain, e.g., near
 $2 \epsilon_{min}$: 
\begin{equation}
R(\omega) \propto  Im \left[\frac{I (\omega)}{1 + 2 V(\pi,\pi) ~I (\omega)}\right],~I (\omega) =  \sum_{k \approx \pi}
 \frac{1}{\omega  - 2 \epsilon (k)}.
\label{a8}
\end{equation}
and $\omega = \omega + i \delta$. 
 Comparing  (\ref{a7}) and (\ref{a8}), we see that there are two differences between $R_0 (\omega) = - Im R_0 (\omega) \propto -Im I(\omega)$
 and $R(\omega)$. First, 
$R(\omega)$ vanishes at $2\epsilon_{min}$ because the divergence in   
 $Im I(\omega)$ in the numerator of (\ref{a8}) is overcompensated by even stronger divergence in the denominator. By the same reason, $R(\omega)$ also
 vanishes at $2\epsilon_{max}$ (and at $2\epsilon_0$ at finite $S$).
Second,   below $2\epsilon_{min}$, $Im I(\omega)$ is small and 
$Re I(\omega)$ is negative and behaves as $-1/\sqrt{2\epsilon_{min} - \omega}$. Then, at some $\omega= \omega_{res} < 2 \epsilon_{min}$, $2 V(\pi,\pi) Re I (\omega_{res}) =0$, and $R(\omega)$ develops a pseudo-resonance. Near $2\epsilon_{max}$, $Re I(\omega)$ is positive, and the resonance does not develop.

In Fig. \ref{fig_2}a we plot $R(\omega)$ obtained by 
solving Eq. (\ref{1}) numerically using Eqs. (\ref{a5}) and (\ref{a6}) for magnon-magnon interaction, and Eq. (\ref{a3}) for the Raman vertex, and by formally extending the quasiclassical formulas to $S=1/2$.
We see that $R(\omega)$ has a sharp peak slightly below $2\epsilon_{min}$.
 The position of the peak and its width somewhat 
depend on the ratio $J_2/J_1$, but for not very small $J_2/J_1$ this dependence is rather moderate. For
 $J_2 = 0.5J_1$, the peak is located near $2J_1$, and its FWHM is $ 0.4 J_1$.
 In between $2\epsilon_{min}$ and $2\epsilon_{max}$, the intensity passes through a maximum, but $R(\omega)$ at the
 maximum is much smaller than at the peak below $2\epsilon_{min}$.
 Like we said, for $S=1/2$, numerical results indicate~\cite{uhrig,uhrig_2}
 that the peak is more broad then in the  quasiclassical analysis extended
 to $S=1/2$. For larger $S$, however, the quasiclassical results should be more accurate.

To conclude, in this paper we argued that the Raman intensity in spin $S$ 
two-leg spin-ladder materials has a pseudo-resonance peak.
 The peak originates from the existence of a local 
minimum in the magnon excitation spectrum, and  is located slightly  below twice the 
 magnon energy at the minimum. The physics leading to the peak is similar to the excitonic scenario for the neutron and Raman resonances in the superconducting state of the cuprates. At large $S$, the peak is quite narrow, its intrinsic width scales as $1/S$. For $S=1/2$, though, the pseudo-resonance may be already 
rather broad. 

 We thank G. Blumberg  and G. S. Uhrig for useful discussions and critical 
 comments.
The research was supported by NSF DMR 0240238 (A.V. Ch)


\begin{thebibliography}{99}
\bibitem{comm} The validity of the large $U$ approach for
$Sr_{14}Cu_{24}O_{41}$ has been  
questioned in recent studies~\protect\cite{FreitasSingh}.
\end{thebibliography}
\end{document}